\documentstyle[sprocl]{article}

\input{psfig}

\def\Journal#1#2#3#4{{#1} {\bf #2}, #3 (#4)}

\def\NAT{\em Nature}
\def\NCA{\em Nuovo Cimento}

\def\NP{\em Nucl. Phys.}
\def\NPA{{\em Nucl. Phys.} A}

\def\PLB{{\em Phys. Lett.}  B}
\def\PRC{{\em Phys. Rev.} C}
\def\PRL{\em Phys. Rev. Lett.}
\def\PRD{{\em Phys. Rev.} D}
\def\PR{\em Phys. Rev.}

\def\ra{\rightarrow}

\def\ko{K^0}

\def\kl{K^{0}_{L}}
\def\al{\alpha}
\def\ab{\overline{\alpha}}
\def\Lamz{\Lambda^0}
\def\Lamb{\overline{\Lambda}^0}
\def\Lam{\Lambda}

\def\Cas{\Xi}
\def\Xim{\Xi^-}
\def\Xib{\overline{\Xi}^+}
\def\ra{\rightarrow}
\def\be{\begin{equation}}
\def\ee{\end{equation}}
\def\bea{\begin{eqnarray}}
\def\eea{\end{eqnarray}}

\begin{document}
\hspace{3.5in}
\begin{tabular}{r}
LBNL-41269\\
Jan 6, 1998\\
\end{tabular}
\vspace{0.5in}

\title{HUNTING FOR DIRECT CP VIOLATION IN STRANGE-BARYON DECAYS 
\footnote
{Invited talk presented at The International Conference on Physics Since
Parity Symmetry Breaking in memory of
Madame C.S. Wu, Nanjing, China, August 16, 1997.}
}

\author{K. B. LUK}

\address{Department of Physics and Lawrence Berkeley National Laboratory,
\\University of California,\\
Berkeley, CA 94720, USA\\E-mail: luk@lbl.gov}

\maketitle\abstracts{
Although CP-symmetry breaking has only been seen in neutral kaon
decays, this mysterious phenomenon is expected to occur elsewhere. 
In this talk, how CP violation can arise in strange-baryon decays is 
briefly reviewed. 
The current status and prospects for searching for such an effect 
will be presented.}

\section{Historical Introduction}

Strange baryons were first discovered in cosmic-ray experiments 
in the late 1940s as particles heavier than the nucleons;~\cite{rb}
hence, they are also known as hyperons.
When Lee and Yang proposed parity (P) nonconservation in 
1956 to explain the famous $\theta$-$\tau$ puzzle, they also suggested 
studying decays of hyperons as further tests of their idea.~\cite{ly} 
Explicitly, they predicted that the distribution of protons in the weak decay 
${\Lamz} \ra p {\pi^-}$ would have a forward-backward asymmetry, 
quantified by the decay parameter $\al_{\Lambda}$, 
with respect to the polarization of $\Lamz$. 
Just a few months after the classic $^{60}$Co experiment performed by 
Madame C.S.~Wu {\it et al}~\cite{wu} was announced, 
the predicted forward-backward asymmetry in $\Lamz$ decay was indeed 
observed.~\cite{aa}

With the fall of parity conservation, invariance of the other discrete 
symmetries, charge conjugation (C) and time reversal (T), 
was also questioned. 
As early as 1957, Lee, Oehme and Yang suggested tests that were sensitive
to T violation in $\ko$ decays.~\cite{loy} 
Within a year, Okubo concluded that breaking of T symmetry would lead to a
difference in the partial 
decay rates between $\Sigma^+$ and $\overline{\Sigma}^-$,~\cite{ok} 
followed by Pais who proposed a test of T violation by comparing the 
magnitude of the $\al$ decay parameters of the $\Lamz$ and 
$\Lamb$ decays.~\cite{ap2}
However, there were arguments suggesting that the combined symmetry of
charge conjugation and parity should be conserved.~\cite{lyl}
These speculations were resolved by the surprising discovery of CP violation 
in $\kl$ decays in 1964~\cite{cp} that led to
a new series of experimental studies and searches for 
CP-symmetry breaking outside of $\kl$.

The early explorations of CP asymmetry in strange-baryon decays were 
limited by statistics.
Due to low energy, it was difficult to produce large samples
of hyperons and anti-hyperons to test the proposal put forward by Okuba.
Pais's idea was not realized experimentally until 1985.~\cite{isr} 
What was done in the 60's and 70's is the determination of the  
$\beta_{\Lambda}$ decay parameter, which is sensitive to T violation if
the final-state interaction is absent,
in the decays of polarized $\Lamz$'s produced in exclusive reactions.~\cite{co} 
However, these attempts were not precise enough to confront any predictions.

Recently, there has been a renewal of interest in hunting for CP asymmetry 
in strange-baryon decays. 
With better instrumentation, it is now
possible to collect a billion hyperon decays with a simple spectrometer 
in a relatively short time.
This will improve the search sensitivity for the CP-odd effect in hyperon
decays by at least two orders of magnitude to a level of $10^{-4}$ or better. 

In the following, theoretical motivation, recent experimental progress
and the prospects for searching for direct CP violation in the nonleptonic
decays of $\Lam$ and charged-$\Xi$ baryons are presented.

\section{Phenomenology}\label{sec:theory}
\subsection{Nonleptonic hyperon decay}
All long-lived spin-1/2 strange baryons decay predominantly into a spin-1/2
baryon and a pion with a change of strangeness $\Delta$S = 1.
\footnote{Except $\Sigma^0$ that decays into $\Lambda^0$
and $\gamma$ by electromagnetic interaction with $\Delta$S = 0.} 
Since parity is not conserved in these weak decays, the orbital angular 
momentum of the final-state particles can be either 0 or 1. 
The corresponding $S$-wave amplitude is parity-violating whereas the $P$-wave
amplitude is parity-conserving. 

In general, a nonleptonic hyperon decay can be described by
the decay rate $\Gamma$ and the decay parameters,
$\al_p$ and $\beta_p$.
In terms of the $S$- and $P$-wave amplitudes, the decay rate is given 
by~\cite{cb}
\vspace{8pt}
\begin{equation}
\Gamma = G_F^{2} m_\pi^4 \frac{p_d (E_d+m_d)}{4\pi m_p}
\big( |S|^2+|P|^2 \big)
\ ,
\label{eq:rate}
\vspace{5pt} 
\end{equation}
where $G_F$ is the Fermi constant,  
$m_{\pi}$ is the mass of the pion, 
$m_p$ is the mass of the hyperon,  
$m_d$ is the mass of the daughter baryon,  
$p_d$ and $E_d$ are the magnitude of the momentum and energy of the baryon 
in the hyperon rest frame respectively.  
The decay parameters, $\al_p$ and $\beta_p$, defined as
\vspace{5pt}
\be
\al_p = \frac{2Re(S^{*}P)}{|S|^2+|P|^2}
\ ,
\quad
\beta_p = \frac{2Im(S^{*}P)}{|S|^2+|P|^2}
\ ,
\label{eq:param}
\vspace{5pt}
\ee
are related to the interference of the $S$- and $P$-wave amplitudes.

In the rest frame of the hyperon, the angular distribution of the 
daughter baryon is 
\vspace{8pt}
\be
\frac{dn}{d \Omega} = \frac{1}{4 \pi} \big( 1 + \al_p 
{\mathbf{P}}_p \cdot \hat{\mathbf{p}}_d \big)
=\frac{1}{4 \pi} \big( 1 + \al_p P_p \cos \theta_d \big)
\ ,
\label{eq:angular}
\vspace{5pt}
\ee
where $\theta_d$ is the angle between the polarization of the parent,
${\mathbf{P}}_p$, and the momentum unit vector of the daughter, 
$\hat{\mathbf{p}}_d$.  
In addition, the polarization of the daughter baryon, ${\mathbf{P}}_d$, 
is related to the polarization of the parent by
\vspace{8pt}
\be
{\mathbf{P}}_d = \frac{(\al_p + {\mathbf{P}}_p \cdot \hat{\mathbf{p}}_d) 
\hat{\mathbf{p}}_d
+ \beta_p {\mathbf{P}}_p \times \hat{\mathbf{p}}_d 
+ \gamma_p \hat{\mathbf{p}}_d \times ({\mathbf{P}}_p \times 
\hat{\mathbf{p}}_d)}
{\big( 1 + \al_p {\mathbf{P}}_p \cdot \hat{\mathbf{p}}_d \big)} 
\ ,
\label{eq:polar}
\vspace{5pt}
\ee
where the decay parameter, $\gamma_p$, given by
\vspace{5pt}
\be
\gamma_p = \frac{|S|^2-|P|^2}
{|S|^2+|P|^2}
\ ,
\label{eq:paramg}
\vspace{5pt}
\ee
is not an independent quantity but is subject to the constraint
$\al^{2}_p+\beta^{2}_p+\gamma^{2}_p=1$.
The decay parameters of the corresponding anti-hyperon decay will
be denoted by $\ab_p$, $\overline{\beta}_p$ and $\overline{\gamma}_p$.

\subsection{CP violation in hyperon decay}
 
\begin{figure}[!htb]
\centerline{\psfig{figure=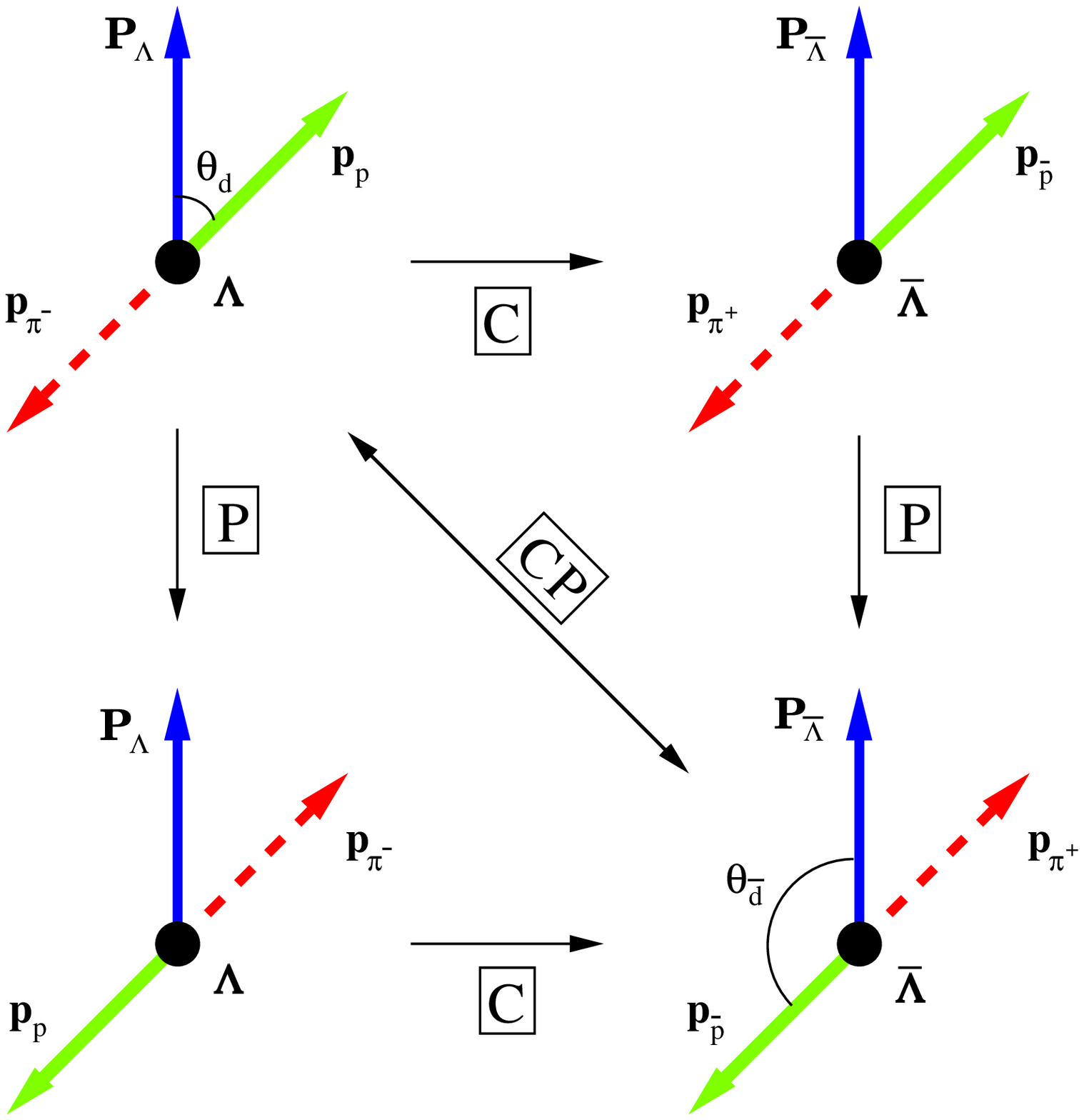,width=4.0in}}
\vspace{15pt}
\caption{C- and P-operations on $\Lam \ra p \pi$ decay. Since 
$\theta_{\overline{d}}$ is
mapped into $\pi-\theta_d$, the decay parameter $\ab_{\Lam}$ is related to
$-\al_{\Lam}$ if CP is conserved.}
\label{fig:cp}
\end{figure}
  
The decay of a strange baryon can be related to that of the corresponding 
anti-strange baryon by CP transformation.
As illustrated in Fig.~\ref{fig:cp} for 
$\Lam \ra p\pi$ decay, under CP operation, 
a daughter baryon emitted in the forward hemisphere defined by the 
polarization of the parent is transformed into an anti-baryon emerged 
in the backward hemisphere.
Therefore, $\ab_p = -\al_p$ under CP transformation. 
In general, if CP is conserved, the decay parameters 
will satisfy the following conditions:
\vspace{8pt}
\be
\quad
\ab_p = -\al_p,
\quad
\overline{\beta}_p = - \beta_p,
\quad
\overline{\gamma}_p = \gamma_p
\ .
\label{eq:abc}
\vspace{5pt}
\ee
These results can be obtained from Eq.~(\ref{eq:polar}) when we impose CP
invariance to the equation.

It is then natural to look for CP-odd effects in strange baryon decays by
comparing the decay parameters of the hyperon and the anti-hyperon. 
The comparison can be realized by defining some CP observables
\vspace{8pt}
\be
\quad
A = \frac {\al+\ab}{\al-\ab} \ ,
\quad
B = \frac {\beta+\overline{\beta}}{\beta-\overline{\beta}} \ .
\label{eq:ABB}
\vspace{5pt}
\ee
Another CP observable relies on the difference in the partial decay rates,
as originally proposed by Okubo,
\vspace{8pt}
\be
\Delta = \frac{\Gamma-\overline{\Gamma}}{\Gamma+\overline{\Gamma}}
\ .
\label{eq:delta}
\vspace{5pt}
\ee

To proceed further, we can write the complex $S$- and $P$-decay amplitudes 
in terms of the moduli, $S_{2{\Delta}I,2I}$ and $P_{2{\Delta}I,2I}$,
the CP-violating weak phases $\phi_{2{\Delta}I,2I}$ and 
the CP-conserving final-state phase shifts $\delta_{2{\Delta}I,2I}$,
where $I$ is the isospin of the final-state and ${\Delta}I$ is the change
of isospin in the decay.
With this parametrization, 
the model-independent CP observables can be expressed approximately as,
for the ${\Lam} \ra p {\pi}$ decay,
\begin{eqnarray}
A_{\Lambda} = -tan(\delta^{P}_{11}-\delta^{S}_1)sin(\phi^{P}_{1}-\phi^{S}_{1}),\qquad \\
B_{\Lambda} = cot(\delta^{P}_{11}-\delta^{S}_1)sin(\phi^{P}_{1}-\phi^{S}_{1}), 
\qquad \quad \\
\Delta_{\Lambda} = \sqrt{2}\frac{S_{33}}{S_{11}}
sin(\delta^{S}_{3}-\delta^{S}_{1})sin(\phi^{S}_{3}-\phi^{S}_{1}),
\ 
\label{eq:lambdaCP}
\vspace{5pt}
\end{eqnarray}
where the strong phases have been measured to be 
$\delta^{P}_{11} = -1.1^{\circ}$,
$\delta^{S}_{1} = 6.0^{\circ}$, and
$\delta^{S}_{3} = -3.8^{\circ}$ with uncertainties of 
about $1^{\circ}$.~\cite{lp} 
For the charged ${\Xi} \ra {\Lam} {\pi}$ decay, 
\footnote{Detailed discussion of CP violation in other strange-baryon decays
can be found in Ref.~14.}
we have
\begin{eqnarray}
A_\Xi = -tan(\delta^{P}_{21}-\delta^{S}_{2})sin(\phi^{P}_{12}-\phi^{S}_{12}),
\\
B_\Xi = cot(\delta^{P}_{21}-\delta^{S}_{2})sin(\phi^{P}_{12}-\phi^{S}_{12}),
\quad \\
\Delta_{\Xi} = 0. 
\ \qquad \qquad \qquad \qquad \qquad \qquad
\label{eq:xiCP}
\vspace{5pt}
\end{eqnarray}
The observable $\Delta_{\Xi}$ is zero because there is only one isospin
state available in the final state.
There is no measurement on the $\Lambda\pi$ re-scattering phases.
Nath and Kumar calculated $\delta^{S}_{2} = -18.7^{\circ}$ and
$\delta^{P}_{21} = -2.7^{\circ}$ whereas Martin got 
$\delta^{P}_{21} = -1.2^{\circ}$.~\cite{nkm}
However, recent calculations argue that both $\delta^{P}_{21}$ and
$\delta^{S}_{2}$ are very small, if not zero.~\cite{lws}

It is interesting to note that we need non-zero CP-violating phase difference
as well as final-state phase shifts to observe direct CP nonconservation in
hyperon decays. 
This is a feature shared by all CP-violating processes.

\subsection{Predictions of CP violation in $\Lambda$ and charged $\Xi$ decays}
Predictions of the CP observables depend on the 
details of the models.~\cite{dep,theory}
In the Kobayashi-Maskawa model, the weak phase is contained in the penguin
diagrams that are responsible for the $\Delta$S = 1 interactions.
In the Weinberg model, the CP asymmetry comes from the exchange of 
charged-Higgs bosons.
Up to now, reliable predictions of CP-odd effects in strange-baryon decays 
have not been not available.
For example, the calculated values of the $\Lambda\pi$ phases 
disagree by an order of 
magnitude, leading to an uncertainty in $A_{\Xi}$ by the same amount.
The situation is further complicated by the fact that 
exact calculation of the hadronic matrix elements 
cannot be implemented in evaluating the weak phases.
Predictions of $A$, $B$, and $\Delta$ have been obtained by using 
different models to calculate the hadronic matrix elements. 
Some predictions of $A$, $B$, and $\Delta$ for the ${\Lam} \ra p {\pi}$
and charged ${\Xi} \ra {\Lam} {\pi}$ decays in the standard model, 
the Weinberg-Higgs model, and the isoconjugate Left-Right symmetric 
model are shown in Table~\ref{tab:predictions}. 
We should note that superweak models do not have $\Delta$S = 1 
CP-odd effects and thus all observables in 
Eqs.~(\ref{eq:ABB}) and~(\ref{eq:delta}) are expected to 
be zero.~\cite{superweak}
\begin{table}[htb]
\begin{center}
\caption{Some predictions of $\Delta$, $A$ and $B$ for 
$\Lam \ra p \pi$ and charged $\Xi \ra \Lam \pi$ decays.
\label{tab:predictions}
}
\vspace{0.4cm}
$\Lam \ra p \pi$ \\
\vspace{0.1cm}
\begin{tabular}{|c|c|c|c|}
\hline
& \raisebox{0pt}[13pt][7pt]{$\Delta_{\Lam}$} 
& \raisebox{0pt}[13pt][7pt]{$A_{\Lam}$} 
& \raisebox{0pt}[13pt][7pt]{$B_{\Lam}$}  \\
\hline
\raisebox{0pt}[13pt][7pt] {CKM}		
&$< 10^{-6}$	
&$(-5$ to $-1) \times 10^{-5}$		
&$(0.6$ to $3) \times 10^{-4}$	\\ \hline
\raisebox{0pt}[13pt][7pt] {Weinberg}	
&$-8 \times 10^{-6}$		
&$-2.5 \times 10^{-5}$	
&$1.6 \times 10^{-3}$		\\ \hline
\raisebox{0pt}[13pt][7pt] {Left-Right}	
&0	
&$(-0.1$ to $6) \times 10^{-4}$	
&$7 \times 10^{-4}$		\\ \hline
\end{tabular}
\end{center}
\begin{center}
\vspace{0.5cm}
\ Charged $\Xi \ra \Lam \pi$ \\
\vspace{0.1cm}
\begin{tabular}{|c|c|c|c|}
\hline
& \raisebox{0pt}[13pt][7pt]{$\Delta_{\Xi}$} 
& \raisebox{0pt}[13pt][7pt]{$A_{\Xi}$} 
& \raisebox{0pt}[13pt][7pt]{$B_{\Xi}$}  \\
\hline
\raisebox{0pt}[13pt][7pt] {CKM}		
&0	
&$(-10$ to $-1) \times 10^{-5}$	
&$(10$ to $1) \times 10^{-3}$	\\ \hline
\raisebox{0pt}[13pt][7pt] {Weinberg}	
&0	
&$-3.2 \times 10^{-4}$	
&$3.8 \times 10^{-3}$		\\ \hline
\raisebox{0pt}[13pt][7pt] {Left-Right}	
&0	
&$(-2.5$ to $6) \times 10^{-5}$	
&$-3.1 \times 10^{-4}$		\\ \hline
\end{tabular}
\end{center}
\end{table}

\section{Searches}
As shown in Table~\ref{tab:predictions}, observable $B$ 
is the most sensitive
probe for finding CP-symmetry breaking in the strange-baryon sector.
To test the predictions of $B$, however, 
will require at least $10^8$ to $10^{11}$ hyperons and anti-hyperons with 
precisely known polarization. 
Furthermore, 
the polarization of the daughter baryon must also be well measured 
by some means, for example by re-scattering,
for determining $\beta$ and $\overline{\beta}$.  
The $\Delta$ observable requires knowing the absolute number of events 
and, in general, it is too small to measure. 
It is unlikely that a meaningful determination of $\Delta$ or  
$B$ will be performed in the near future. 
The best observable for finding CP violation in 
strange-baryon decays is $A$.
Yet, measuring $A$ still asks for decay samples with 
well determined hyperon polarization.

\subsection{What have been done}
There were three attempts to search for CP nonconservation in $\Lam$ decay
by experiments R608 at ISR,~\cite{isr} 
DM2 at Orsay,~\cite{dm2} and PS185 at LEAR.~\cite{ps185a,ps185b}
However, none of these experiments was designed specifically for studying
CP symmetry in strange-baryon decays. 
Based on a limited number of polarized ${\Lamz} \ra p {\pi^-}$ and 
${\Lamb} \ra {\overline p}{\pi^+}$ decays, the best result for $A_{\Lam}$
is -0.013$\pm$0.022.~\cite{ps185b} 
We will summarize below the highlights of these three measurements.

\vspace{0.35in}
\noindent
\textbf{R608 at ISR}

Polarized $\Lamz$'s and $\Lamb$'s were 
produced in the forward beam 
fragmentation region in the inclusive reactions $pp \ra {\Lamz}+X$
and ${\overline p}p \ra {\Lamb}+X$, respectively.
At the sensitivity of this experiment,
C invariance in strong interactions is valid, and the
polarization of $\Lamz$ produced in $pp \ra {\Lamz}+X$ is identical to that
of $\Lamb$ created in ${\overline p}p \ra {\Lamb}+X$.  
Based on 17,028 ${\Lamz} \ra p {\pi^-}$ and 9,553
${\Lamb} \ra {\overline p}{\pi^+}$ events, 
the ratio $\al_{\Lam}/\ab_{\Lam}$ was measured to be $-1.04 \pm 0.29$,
which is equivalent to $A_{\Lam}$ = $-0.02 \pm 0.14$.

\vspace{0.35in}
\noindent
\textbf{DM2 at Orsay}

A total of 1,847 ${\Lamz}{\Lamb}$ pairs from the decays of 
J/${\Psi}$'s that were produced in unpolarized $e^+ e^-$ collisions 
was used to study the differential cross section
\begin{eqnarray}
\frac{d{\sigma}}{d{\cos{\theta_{\Lam}}}d{\Omega_p}d{\Omega_{\overline{p}}}}
\propto
2{\Big(1-\frac{p^{2}_{\Lam}}{E^{2}_{\Lam}}{\sin^{2}{\theta_{\Lam}}}\Big)}
{\big[1-{\al_{\Lam}}{\ab_{\Lam}}({\mathbf{p}}\cdot{\mathbf{n}})
({\mathbf{\overline{p}}}\cdot{\mathbf{n}})\big]}
\nonumber\\ 
+\frac{p^{2}_{\Lam}}{E^{2}_{\Lam}}{\sin^{2}{\theta_{\Lam}}}
\Big\{1-{\al_{\Lam}}{\ab_{\Lam}}{\big[{\mathbf{p}\cdot{\mathbf{\overline{p}}}}
-2({\mathbf{p}\cdot{\mathbf{x}}})
({\mathbf{\overline{p}}}\cdot{\mathbf{x}})\big] } \Big\}
\ ,
\label{eq:xsect}
\end{eqnarray}
where $\theta_{\Lam}$ is the emission angle of $\Lamz$ with respect to 
the $e^+$ beam direction, $p_{\Lam}$ and $E_{\Lam}$
are the momentum and energy of $\Lamz$ in the J/$\Psi$ decay, 
$\mathbf{p}$ and $\mathbf{\overline{p}}$ are the p and $\overline{p}$ 
momentum in the $\Lamz$ and $\Lamb$ rest frame respectively, 
$\mathbf{x}$ is normal to the plane formed by the $\Lamz$ momentum
and the beam axis, and $\mathbf{n}$ is a unit vector that 
suppresses the spin state of zero in the J/$\Psi$ decay and is 
degenerate with the beam axis at $\theta_{\Lam}$ = 
$0^{\circ}$ and $90^{\circ}$.  
By fixing the decay parameter $\al_{\Lam}$ at the canonical 
value of 0.642,~\cite{pdg} the observed distribution for 
$\mathbf{p}{\cdot}{\overline{\mathbf{p}}}$ was fitted to 
the differential cross section in Eq.~(\ref{eq:xsect}) 
by varying $\ab_{\Lam}$.  
The least squares minimization yielded $\ab_{\Lam}$ = $-0.63 \pm 0.13$. 
Consequently, $A_{\Lam}$ was determined to be $0.01 \pm 0.10$.

\vspace{0.35in}
\noindent
\textbf{PS185 at LEAR}

In this experiment, polarized ${\Lamb}{\Lamz}$ pairs were produced in 
exclusive reaction $\overline{p}p \ra {\Lamb}{\Lamz}$ at threshold. 
Again, conservation of charge conjugation in the production process 
guarantees the polarization of $\Lamz$ and $\Lamb$ to be the same.  
From the ratio of ${\al_{\Lam}}P_{\Lam}$
to ${\ab_{\Lam}}{\overline{P}}_{\Lam}$,
a value of $0.013 \pm 0.022$ for $A_{\Lam}$ was obtained recently.

\subsection{Current Search: HyperCP (Experiment 871) at Fermilab}
\begin{figure}[hbt]
\centerline{\psfig{figure=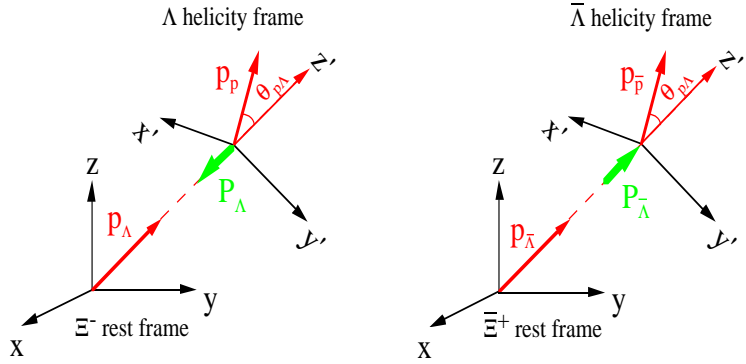,height=2.0in,width=4.0in}}
\vspace{15pt}
\caption{Helicity frame used for determining the angular 
distribution of the decay baryon. Since the $\al$ decay
parameters change sign under CP transformation the polarizations of $\Lamz$
and $\Lamb$ are equal and opposite. The orientation of the
helicity frame is not fixed in space but changes from event to event.}
\label{fig:alfalf}
\end{figure}

HyperCP prepares polarized $\Lamz$ and $\Lamb$ samples from the
decay of $\Xim$ and $\Xib$ hyperons.
When an unpolarized $\Xim$ decays, from Eq.~(\ref{eq:polar}), 
the daughter $\Lamz$ acquires longitudinal polarization which is given by
\vspace{5pt}
\be
{\mathbf{P}}_\Lam = \al_{\Cas} \hat{\mathbf{p}}_{\Lam} \ .
\label{eq:helicity}
\vspace{5pt}
\ee
In other words, the $\Lamz$ polarization is \emph{absolutely} determined by 
the decay parameter $\al_{\Cas}$.
In this case, the angular distribution of the decay proton along 
$\hat{\mathbf{p}}_{\Lam}$ in the helicity frame of $\Lamz$, as shown in 
Fig.~\ref{fig:alfalf}, is
\vspace{8pt}
\be
\frac{dn}{d \cos{\theta_{p\Lam}}} =\frac{1}{2} \big(1+\al_{\Lam} 
\al_{\Xi} \cos \theta_{p\Lam} \big)       
\ .
\label{eq:cosplam}
\vspace{5pt}
\ee
Similarly, the decay distribution of $\overline{p}$ in the $\Lamb$
helicity frame has a slope of $\ab_{\Lam} \ab_{\Xi}$.
As discussed in Sec.~\ref{sec:theory}, both decay parameters change
sign under CP transformation.
Thus, any difference in the slopes of the angular distributions
between $p$ and $\overline{p}$ will signal breaking of CP invariance in
the $\Cas \ra \Lam \pi$, $\Lam \ra p\pi$ decay sequence.
The degree of direct CP violation can be quantified by an observable
\vspace{5pt}
\be
A_{\Lam \Xi} = \frac{\al_{\Lam} \al_{\Xi}-\ab_{\Lam} \ab_{\Xi}}
{\al_{\Lam} \al_{\Xi}+\ab_{\Lam} \ab_{\Xi}}
\simeq A_{\Lam}+A_{\Xi} \ .
\label{eq:alfa}
\vspace{5pt}
\ee
That is, the measurement is sensitive to any CP-odd effect
in the $\Lam$ and charged $\Cas$ decay.

A unique feature of the helicity frame defined by the polarization of $\Lamz$ 
in Eq.~(\ref{eq:helicity}) is that its orientation varies from 
event to event with 
respect to the fixed laboratory coordinate system. 
As a result, any imperfection, temporal variation and systematic 
effects that are localized 
in space are mapped into a wide range of $\cos \theta_{p\Lam}$. 
Hence the systematic biases are highly diluted in the helicity 
frame and have little impact on the determination of $A_{\Lam \Xi}$.
  
\begin{figure}[tb]
\centerline{\psfig{figure=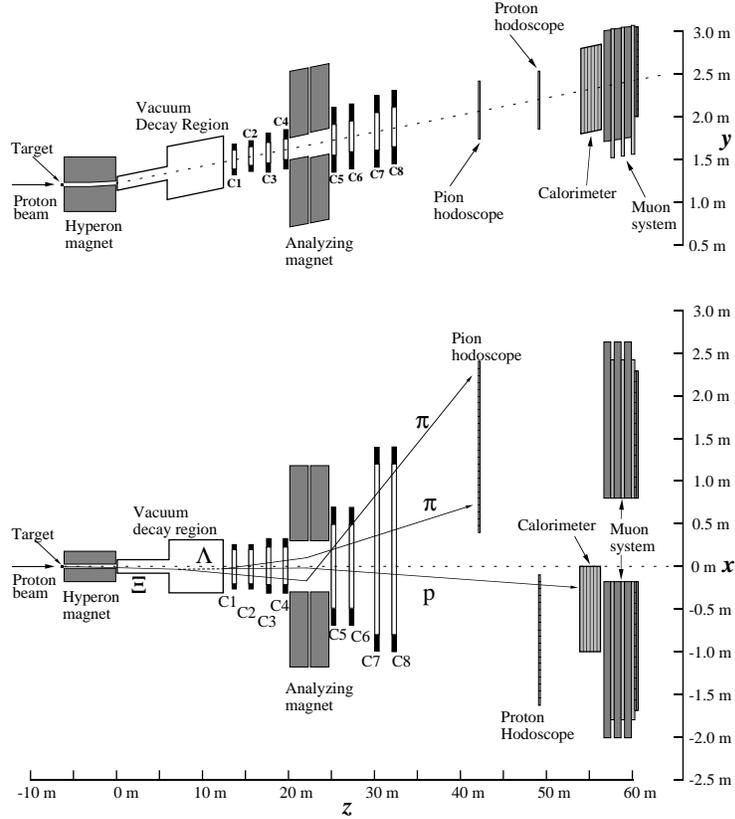,width=4.0in}}
\vspace{25pt}
\caption{Schematic drawing of HyperCP spectrometer.}
\label{fig:e871spectrometer}
\end{figure}

The goal of HyperCP is to reach a sensitivity of $10^{-4}$ in $A_{\Lam \Xi}$.
The experiment is taking its first data at Fermilab.
The plan and elevation views of the HYperCP spectrometer are shown 
in Fig.~\ref{fig:e871spectrometer}.
The spectrometer is kept very simple to minimize any potential bias
to the measurement.
An $800~GeV/c$ proton beam, with a typical intensity of $1.5 \times 10^{11}$
protons per $20~s$, strikes either a $6~cm$-long or a $2~cm$-long 
copper target at a mean angle of $0^{\circ}$.
The longer target is used for producing $\Xim$ hyperons.
Two different targets are employed to ensure the singles rate 
in the spectrometer is comparable between the $\Xim$ and the $\Xib$ runs.
Parity invariance in strong interactions guarantees the polarization of the
produced particles to be zero.
The secondary charged beam is momentum and charge selected
by a curved collimator located inside a $6~m$-long dipole magnet.
The channel has a limited acceptance in transverse momentum, further ensuring
any residual $\Xi$ production polarization to be small.
The accepted $\Xi$ momentum is between $120~GeV/c$ and $240~GeV/c$, with a
mean value of about $167~GeV/c$.
Behind the collimator is a vacuum decay region where most of $\Cas$ 
baryons decay.
The charged particles from $\Cas$ and $\Lam$ decays are detected with
a magnetic spectrometer made up of multiwire proportional chambers with
high-rate capability. 
After the momentum-analyzing magnet, the daughter proton  
is deflected toward the proton hodoscope whereas the pions from $\Xim$ 
and $\Lamz$ decays go toward the pion hodoscope.
These decay products are well separated from the secondary charged beam
by the time they reach the trigger hodoscopes.
The magnetic fields of all spectrometer magnets are monitored 
with high-precision Hall probes.
	
A simple, yet selective trigger for collecting $\Xi$ and $\Lam$
decays is formed by requiring a time coincidence of signals from 
the proton and pion hodoscopes as well as the hadron calorimeter.
The calorimeter is used to reduce the trigger rate which is 
otherwise dominated by secondary 
interactions of the charged beam with material in the spectrometer. 
In addition, events with background muons are highly suppressed by the
energy requirement of the calorimeter in the trigger. 

By reversing the polarities of the magnets in the spectrometer
and switching to the shorter target,
$\Xib$ decays are collected with the identical CP-invariant trigger. 
To minimize bias due to temporal variation in the experiment,
the $\Xim$ and $\Xib$ runs are cycled at least once a day.
 
The data acquisition system, based on multiple VME crates, 
is designed to handle a maximum trigger rate of $100,000~Hz$,
with a maximum throughput of about $17~Mb/s$.~\cite{daq} 
At the nominal proton intensity, the total trigger rate is approximately
$75,000~Hz$, with a mean event size of about $550~bytes$.
 
\begin{figure}[tb]
\centerline{\psfig{figure=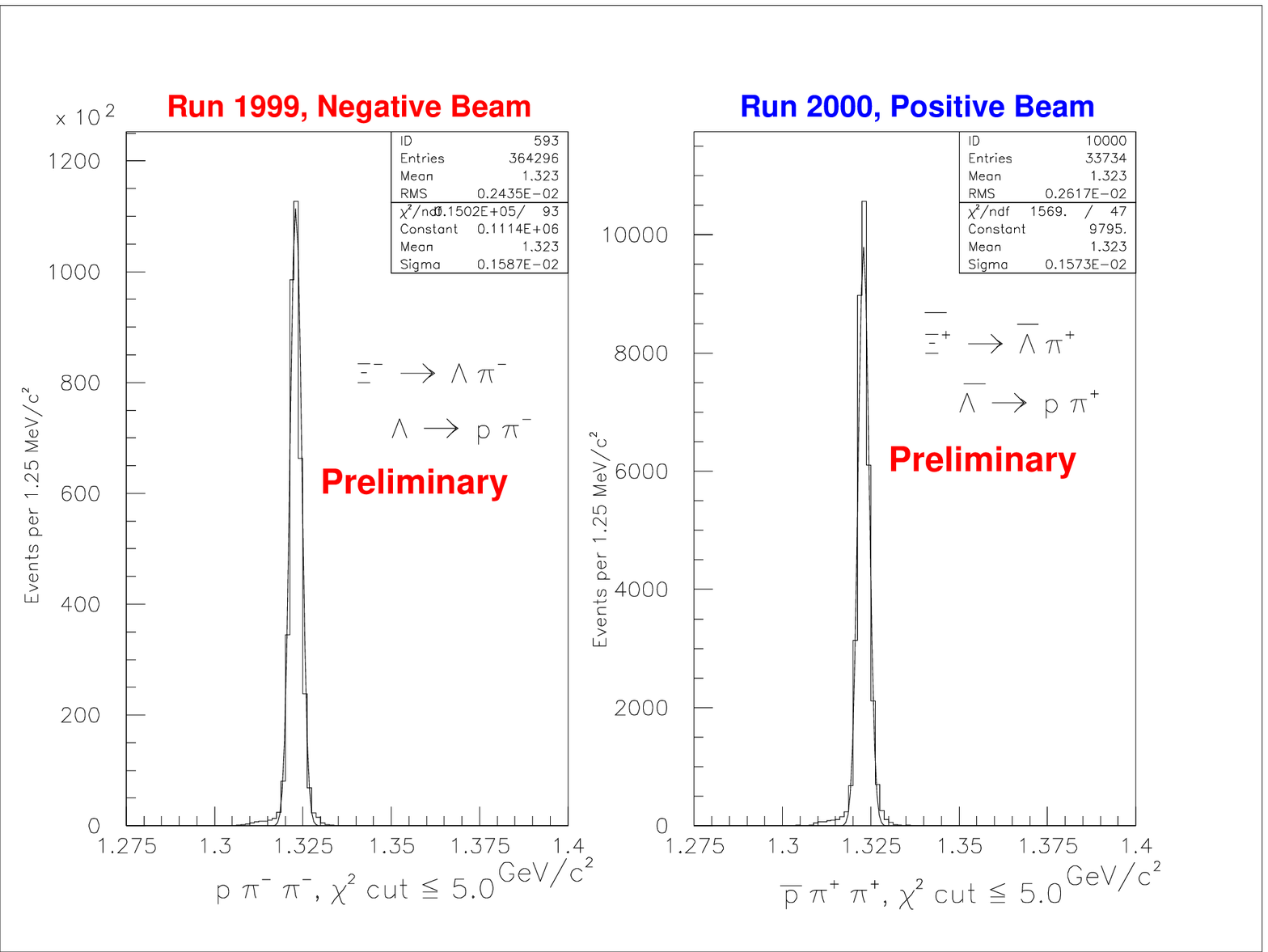,width=4.0in}}
\vspace{25pt}
\caption{$\Lam$-$\pi$ invariant mass distributions from the $\Xim$ and
$\Xib$ runs. These events pass geometric and kinematic fits. 
A cut on the quality of the kinematic fit has been made. 
The mass resolution is $1.6~MeV/c^2$.}
\label{fig:mass}
\end{figure}

The 1996-1997 run will end on September 5th. 
HyperCP will have recorded $7.5 \times 10^{9}$ triggers, 
of which $1.5 \times 10^{9}$
are $\Xim$ triggers and $2.5 \times 10^{9}$ are $\Xib$ triggers.
Preliminary analysis indicates that this will yield approximately
$9 \times 10^{8}$ $\Xim$ and $2.5 \times 10^{8}$ $\Xib$ decays.
The anticipated statistical uncertainty in $A_{\Lam \Cas}$ is
about $2 \times 10^{-4}$ in the first run.
Preliminary $\Lam \pi$ invariant-mass distributions for a 
small number of events 
from the $\Cas$ trigger are shown in Fig.~\ref{fig:mass}.
An excellent mass resolution of $1.6~MeV/c^2$ has already been achieved.
Notice the similarity of the mass peaks, indicating that the running condition 
is stable between runs.

\section{Prospects}
After the commissioning of the Main Injector at Fermilab, HyperCP will have a
second run in 1999.
With minor improvements to the spectrometer and simple upgrades to the DAQ
system, HyperCP should be able to take data at higher beam intensity.
Combining the data sets of the two runs should yield an
uncertainty in $A_{\Lam\Xi}$ of $10^{-4}$ or better.

It has been suggested that the tau-charm factory can provide very clean 
and systematic free conditions for investigating CP violation 
in $\Lam$ and $\Xi$ decays.~\cite{tauf} 
When a tau-charm factory is operated at the $J/\psi$ peak, 
the production cross section of $J/\psi$ strongly depends on 
the beam-energy spread.  
By using monochromators the energy spread can be reduced 
from $1.3~MeV$ to $0.1~MeV$ and the expected luminosity is 
about $4 \times 10^{32}~ cm^{-2}s^{-1}$.  
In $10^7s$ of running, 
the number of $\Lamz\Lamb$ and $\Xi\overline{\Xi}^+$ pairs 
coming from $J/\psi$ decays is 
estimated to be $1.1 \times 10^8$ and $1.4 \times 10^8$ respectively. 
	
%
What CP observables that the tau-charm factory can measure depends on 
the polarization of the colliding beam.
Specifically the error in $A_{\Lam}$ and in $A_{\Xi}$ that 
can be achieved in $10^7s$ is about $10^{-3}$.

\section{Conclusion}
Although it was proposed right after the discovery of parity nonconservation
in weak interactions and almost a decade before the observation of 
CP violation in $\kl$ decay, the study of CP symmetry in 
strange baryon decays did not flourish.
The current limit of testing CP invariance in hyperon decays is 
only at the $10^{-2}$ level, which is at least two orders
of magnitude away from most theoretical predictions.  
There is no dedicated experiment with better precision in the near future 
except HyperCP at Fermilab that will reach a sensitivity of 
$2 \times 10^{-4}$ in the 1997 run.  
The outcome of HyperCP could play an important role in 
defining the future program of studying CP violation in strange baryon decays.

\section*{Acknowledgments}
I would like to express my gratitude to the organizers, especially Professor
Feng Wang, for inviting me to attend this conference.
Although I have never met Madame C.S. Wu, I have always greatly 
respected her.
I also thank Sandip Pakvasa and Mahiko Suzuki for many stimulating 
discussions about the subject, and my colleagues on HyperCP who 
help me to realize my dream.
I am indebted to Craig Dukes for his careful reading of the manucript. 
This work was supported by the Director, Office of Energy Research, Office of
High Energy and Nuclear Physics, Division of High Energy Physics of the U.S. 
Department of Energy under Contract DE-AC03-76SF00098.

\end{document}